\documentclass{IEEEtran}
\usepackage{url}
\usepackage{cite}
\usepackage{amsmath,amssymb,amsfonts}
\usepackage{graphicx}
\usepackage{textcomp,nicefrac}
\usepackage{siunitx}
\graphicspath{{img/}}
\usepackage[T1]{fontenc}

\begin{document}
\bstctlcite{IEEEexample:BSTcontrol}

\title{Synchronization of PLC Subsystems with the Timing Distribution System at the European XFEL}
\author{Dmytro Levit, Bruno Fernandes, Peter Zalden, Tobias Freyermuth, Navid Mashayekh, and Patrick Gessler
    \thanks{Manuscript received 6 July 2026.}
    
    \thanks{D. Levit, B. Fernandes, P. Zalden, T. Freyermuth, N. Mashayekh, and P. Gessler are with the European XFEL, 22869 Schenefeld, Germany (e-mail: dmytro.levit@xfel.eu).}
}

\maketitle

\begin{abstract}

The European XFEL generates bursts of up to 4096 ultra-short X-ray flashes with a spacing of only 222\,ns every 100\,ms.
These flashes are produced in the linear electron accelerator by undulators and guided 1 kilometer in vacuum pipes until they reach the scientific experiments.
Accurate synchronization and the availability of information about all configurable pulses within a single burst, which can vary every 10 Hz cycle, enable the full potential of control, monitoring, and measurement in terms of accuracy and utilization.
The MicroTCA-based timing system provides all the necessary capabilities.
However, the industrial programmable logic controllers (PLCs), which manage most of the control electronics of the beamlines and experiments, are only synchronized on a millisecond-level through UART and NTP interfaces.
This paper presents a new development based on an FPGA SoC EtherCAT solution in MicroTCA that implements an interface to the PLC using distributed clocks to achieve synchronization on nanosecond-level and provide beam-related information in a deterministic manner.

\end{abstract}

\pagestyle{empty}


\section{Introduction}
\label{sec:introduction}
\begin{figure}
    \includegraphics[width=0.45\textwidth]{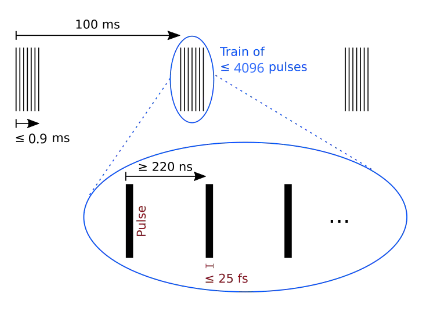}
    \caption{Pulse structure of the beam at the European XFEL}
    \label{fig:euxfel_pulse_structure}
\end{figure}

\IEEEPARstart{T}{he} European X-Ray Free-Electron Laser\,(XFEL) is a research facility in Schenefeld, Germany, that provides high-brilliance X-ray beams to seven scientific instruments\,\cite{euxfel}.
A 3\,km long superconducting linear accelerator, originating at DESY\,(Deutsches Elektronen-Synchrotron) in Hamburg, accelerates electrons to 17.5\,GeV and delivers the beam to three beamlines at the facility, where the undulators use the self-amplifying spontaneous emission effect\,\cite{kondratenko1980generating} to generate coherent X-ray light from the electrons.
Figure\,\ref{fig:euxfel_pulse_structure} shows the pulse structure of the beam.
The accelerator delivers beam in a 900\,µs train within a 100\,ms accelerator cycle.
A train can contain up to 4096 pulses, each with a duration of less then 25\,fs, while adjacent pulses are separated by 222\,ns.
The instruments use the resulting X-ray light with wavelengths between 0.50\,\AA\,and 47\,\AA\,and ultra-short pulse duration to study the dynamics of chemical reactions in organic and inorganic samples, as well as fundamental light-matter interactions.

The beamlines in the facility and the instruments use industrial automation solutions based on the Programmable Logic Controllers\,(PLCs) and bus terminals.
Until now, we have configured and operated these solutions synchronized via the Network Time Protocol\,(NTP) which is far too imprecise for scheduling events relative to the arrival time of X-ray pulses.

This paper describes the synchronization of the PLC system with the facility timing distribution system with nanosecond-level precision using the Distributed Clocks\,(DC)\,\cite{ethercat_dc} mechanism of the EtherCAT fieldbus, thereby enabling hardware control based on beam timing and bunch energy.
Better synchronization opens opportunities to improve system control and enable more sophisticated measurements with the PLC system.

The paper has the following structure: section\,\ref{sec:plc_ethercat} describes the PLC subsystem and the EtherCAT fieldbus, section\,\ref{sec:timing_overview} gives the overview of the timing information distribution system at the European XFEL, section\,\ref{sec:synchronization_bridge} describes the hardware board and the FPGA-based synchronization mechanism, section\,\ref{sec:measurements} describes the test setup and measurement of system synchronization, and section\,\ref{sec:outlook} gives an outlook for possible applications of the system at the European XFEL.

\section{PLC System at the European XFEL}
\label{sec:plc_ethercat}

There are more than 7000 PLC terminals at the European XFEL.
We use the PLC system to control electro-mechanical components of the scientific instruments, measure their position, control and monitor vacuum components, and in safety applications to protect the equipment.

To simplify maintenance and speed up development cycles, the PLC team at the European XFEL decided very early on to use the PLC terminals and software from Beckhoff GmbH exclusively\,\cite{plc}.
The Beckhoff PLC software and terminals natively use EtherCAT\,\cite{doi:10.1049/cce:20040104} fieldbus for communication.

EtherCAT is a real-time Ethernet communication protocol based on the 100Base-TX Ethernet standard\,\cite{ieee8023u1995} and later revised to support faster Ethernet standards.
The fieldbus always requires a master device, which initiates the transaction.
This is usually a software running on an industrial PC.
All slave devices are connected in a loop, a star, or a mixed topology.
In the loop topology, the devices are connected in the daisy chain and EtherCAT frames traverse all devices, exchanging data on the fly as they pass through each device.
In the star topology, the devices are connected directly to the master.
The mixed topology is the mixture of the star and the loop topologies.

The distinctive feature of EtherCAT, the Distributed Clocks, is its deterministic transport latency and drift compensation which corrects differences in clock frequencies between endpoints.
The DC makes it possible for the PLC software to synchronize system time counters in the terminals with a precision below 100\,ns\,\cite{ethercat_dc}.
All terminals in the loop get a common time reference frame and, therefore, can reliably exchange the timestamp of events from different devices in the same loop.

EtherCAT implementations follow the OSI model\,\cite{osi_model}.
The physical layer is implemented in Ethernet PHY ASICs supporting 100Base-TX.
The data link layer requires real-time data processing and is implemented in an ASIC or an FPGA.
The application layer can be implemented in software or in firmware.
In this work, we implemented both the data-link and the application layer in FPGA firmware.


\section{Timing Distribution System}
\label{sec:timing_overview}

The timing distribution system is a hierarchical system used for control of the electron accelerator\,\cite{euxfel_timing}.
The timing system distributes accelerator clock and beam configuration through the network of timing boards to detectors and data acquisition hardware.

The timing distribution system of the facility is based on the x2timer\,\cite{x2timer} boards in the Advanced Mezzanine Card\,(AMC) form-factor.
The boards build a multi-level timing distribution system using 1.3\,Gbps serial optical links with drift and offset compensation to distribute beam information.
The boards provide the following information about the accelerator cycle:
\begin{itemize}
    \item the unique cycle ID (train ID),
    \item the precise timing for the start of the accelerator cycle,
    \item the full table of pulses, and
    \item the beam mode, which provides the mask of the active segments in the accelerator.
\end{itemize}
The pulse table can change from cycle to cycle and consists of entries containing pulse information: pulse position, pulse charge, and destination beamline.
The final-level timing board decodes this information and provides it over a 108\,Mbps Universal Synchronous and Asynchronous Receiver-Transmitter link, USART, on the backplane of the Micro Telecommunications Computing Architecture\,(uTCA) crate to other boards for synchronization.

In addition, the timing system provide a reduced subset of the timing information over a 115\,kbps USART link to the PLC system.
The reduced subset of the timing information includes only the train ID and the beam mode.
Receiving this information signals the start of a new accelerator cycle to the PLC system which can use this as a trigger for its actions.

The low data rate limits the amount and timing precision of information and makes synchronous operation of the PLC system with the accelerator impractical for fast control.
At the moment, changing or measuring the position of the mechanical equipment requires disruption to data acquisition as we have large uncertainty about the read-out time relative to the arrival time of X-ray pulses.
The following section presents our approach to bridge the timing distribution and the PLC systems, and make precise information about the beam and its timing available to the PLC terminals.

\section{PLC Synchronization with the Timing Distribution System}
\label{sec:synchronization_bridge}

\begin{figure}
    \includegraphics[width=0.45\textwidth]{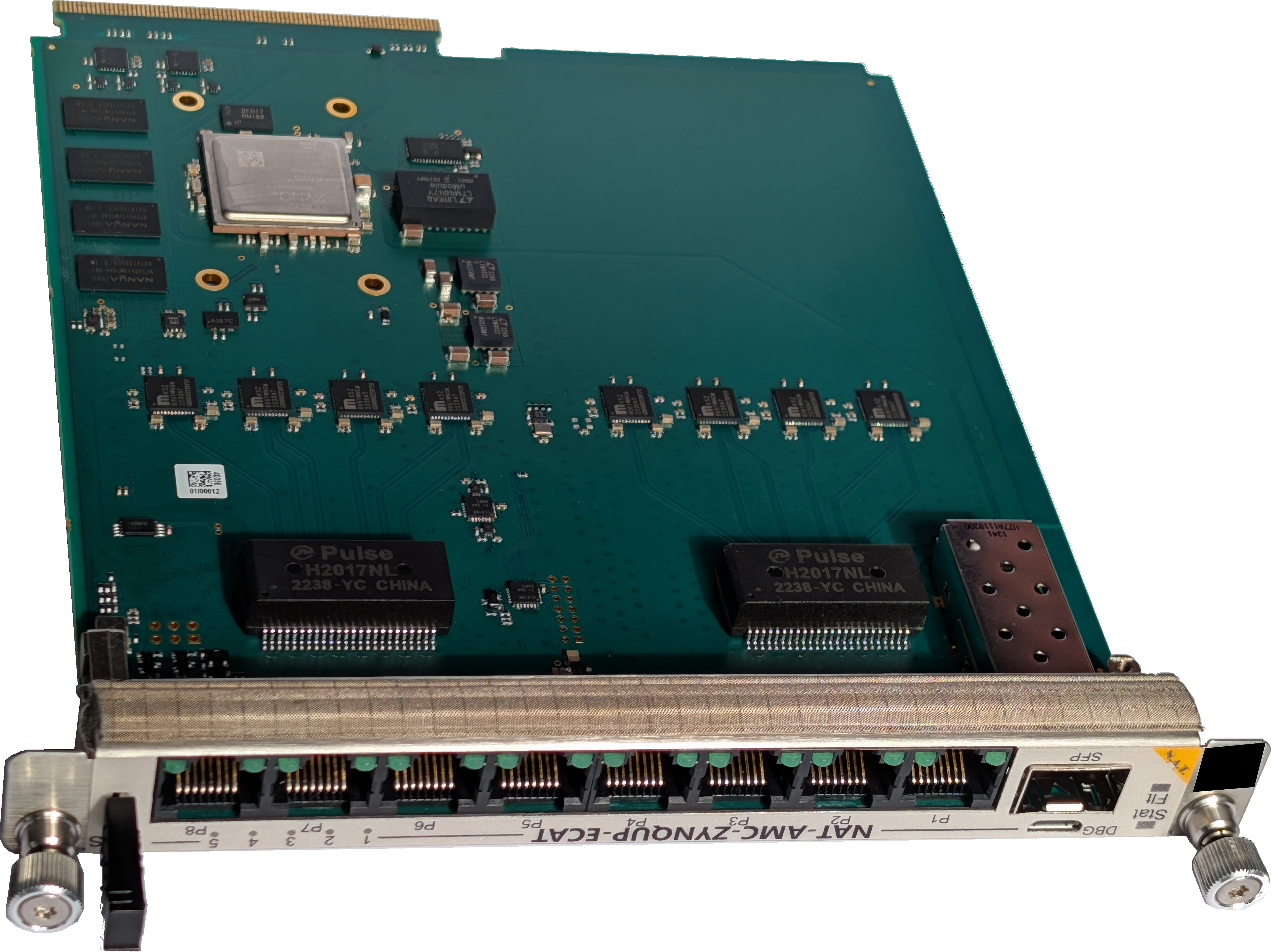}
    \caption{The Zynq UltraScale+ MPSoC board in the uTCA form factor we use for interfacing the timing distribution system to the PLC system over EtherCAT interface.}
    \label{fig:ecatnat}
\end{figure}

\subsection{Synchronization Principles}

The PLC and the timing systems use two independent time reference frames.
Therefore, we must find a common reference frame for both systems or convert the time between the reference frames.
Because the timing metadata carries an NTP timestamp of the start of cycle pulse, we evaluated passing this timestamp to the PLC system, but decided against it because NTP does not provide sufficient precision.

We decided to use the backplane trigger signal which is generated by the x2timer instead.
The x2timer generates this signal with picosecond precision and the FPGA measures its arrival time in the reference frame of the EtherCAT distributed clocks in the clock domain of the 100\,MHz EtherCAT clock.
Thus, the resulting timestamp has a precision of 10\,ns and, because the time counters in the endpoints are synchronized via the DC mechanism, the timestamp maintains a fixed relation to the beam and can be used to calculate all further actions.

\begin{figure*}
    \includegraphics[width=0.95\textwidth]{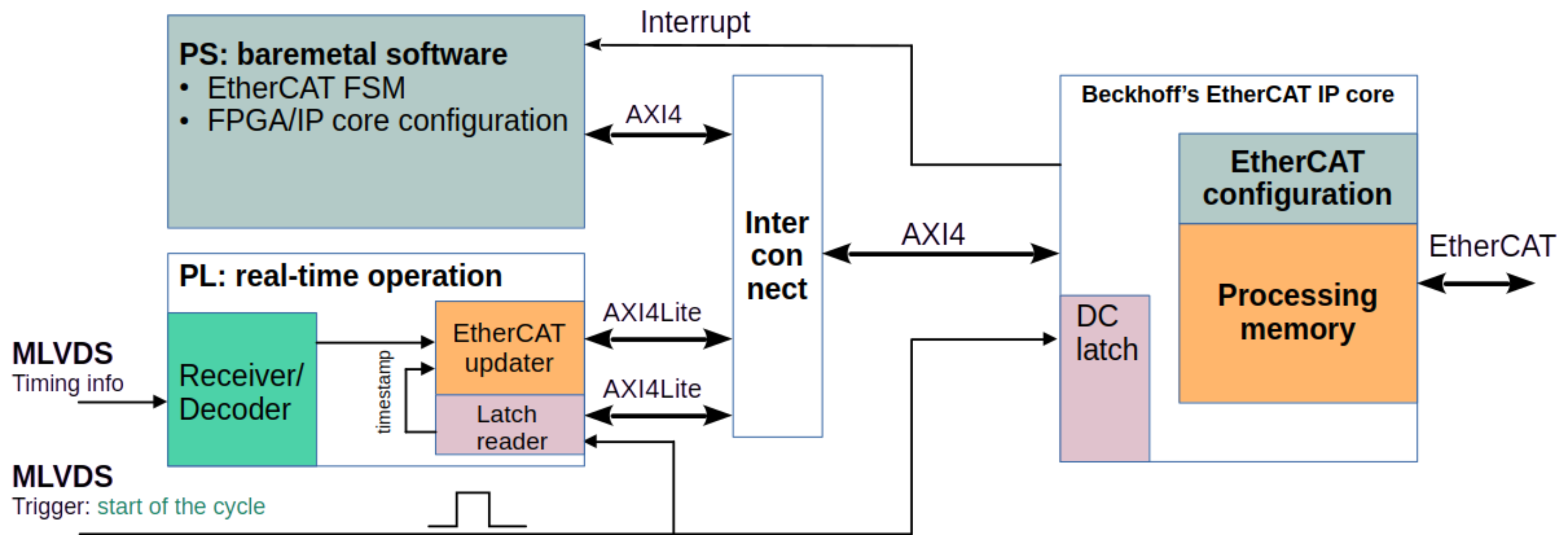}
    \caption{Layout of the firmware for the communication bridge. The firmware consists of the Beckhoff IP core for EtherCAT communication, baremetal software in the processing system for non-real-time control, and the real-time data processing in the programmable logic.}
    \label{fig:firmware_layout}
\end{figure*}

\subsection{Hardware}

At the European XFEL, we maintain uTCA-based infrastructure consisting of timing distribution systems and analog-digital converter boards.
Therefore, we require the bridge hardware to be in the uTCA form factor.
Further requirements include EtherCAT-compatible network interfaces, and an FPGA-based processing system for real-time data processing and communication.

As there were no boards on the market that satisfied our requirements, we collaborated with the N.A.T. GmbH to produce a board in the uTCA form factor with the capabilities listed above.
Figure\,\ref{fig:ecatnat} shows the picture of the produced NAT-AMC-ZYNQUP-ECAT board\,\cite{nat}.

The board is built around the Zynq ZynqUltraScale+ Multiprocessor System-on-Chip\,(MPSoC) XCZU2CG.
The MPSoC is connected to the backplane on the AMC connector, from where it receives timing metadata and the precise start-of-cycle pulse from the x2timer over the Multipoint Low-Voltage Differential Signaling\,(M-LVDS) lines, and to 8 RJ45 ports on the front panel, each capable of 100\,Mbps Ethernet communication.

To attach a board to the EtherCAT loop, we need a minimum of two RJ45 ports.
Therefore, the board can be used to supply timing information to 4 PLC loops in parallel.

\subsection{Firmware}

Figure\,\ref{fig:firmware_layout} shows the layout of the firmware designed for the communication bridge.
The firmware consists of the EtherCAT IP core\,\cite{beckhoff_ipcore} for communication with EtherCAT endpoints, real-time data processing block in the programmable logic, and the baremetal software for non-realtime configuration in the processing system of the MPSoC.

To communicate with EtherCAT endpoints, the device must implement an EtherCAT slave controller in the data link layer and an application layer.
A typical EtherCAT terminal usually uses an ASIC as an EtherCAT slave controller.
In this paper, we use the EtherCAT IP core v3.00k from Beckhoff to implement the EtherCAT slave controller.
The IP core takes over the steering of the Ethernet PHY chips, EtherCAT communication, and synchronization and drift compensation for the distributed clocks.

For data exchange with the data processing logic, the IP core provides memory-based access through the Advanced eXtensible Interface 4\,(AXI4) interface.
The memory is subdivided into two regions: the EtherCAT {\bf configuration} memory and the {\bf processing} memory.
The EtherCAT configuration memory is the unbuffered memory accessible to both the PLC master and the processing logic.
It contains the registers common to all EtherCAT slave controllers that are necessary for communication over the fieldbus.
The configuration memory plays an essential role in establishing communication and identification of the device.
The processing memory is a memory available for exchanging information between the PLC master and the application layer on the device.

As an additional feature, the IP core provides a distributed clock latch function which latches the timestamp of the rising edge of the received signal and stores it in the EtherCAT configuration memory.

The baremetal software configures the on-board hardware, monitors and reacts to the interrupts from the IP core to configure the device ID and establish the EtherCAT communication.
After communication is established, the software is responsible for changing the states of the EtherCAT finite-state machine in response to the PLC master request, and acknowledging the state transition.

The application layer in the programmable logic receives timing metadata from the x2timer and decodes them.
The logic also implements two AXI4Lite master modules.
One master reads the timestamp from the EtherCAT configuration memory.
Another master reads the timing metadata and the timestamp, and writes them to the processing memory of the IP core.
When the last memory address in the IP core is accessed, the IP core marks the buffer as finished, and the written information is available to be accessed by the PLC master.

The AXI4Lite master that writes data to the processing memory and the PLC master must use synchronized mapping of the information stored in the processing memory.
Unsynchronized mapping can lead to delays in accessing data by the PLC rendering the data out of date or causing data misalignment.
To avoid this situation, we define memory content in a SystemRDL\,\cite{systemrdl} file and use this file to generate the AXI4Lite master and the EtherCAT SubDevice Information\,(ESI) XML file for the PLC software.
To use the PeakRDL framework\,\cite{peakrdl} for the generation of the master module and the XML file, we designed the PeakRDL plugins and offer them as open-source software\,\cite{peakrdl-axil-vhdl},\,\cite{peakrdl-ethercat-xml}.

\section{Measurement of Synchronization on a Test Stand}
\label{sec:measurements}


\begin{figure}
    \includegraphics[width=0.5\textwidth]{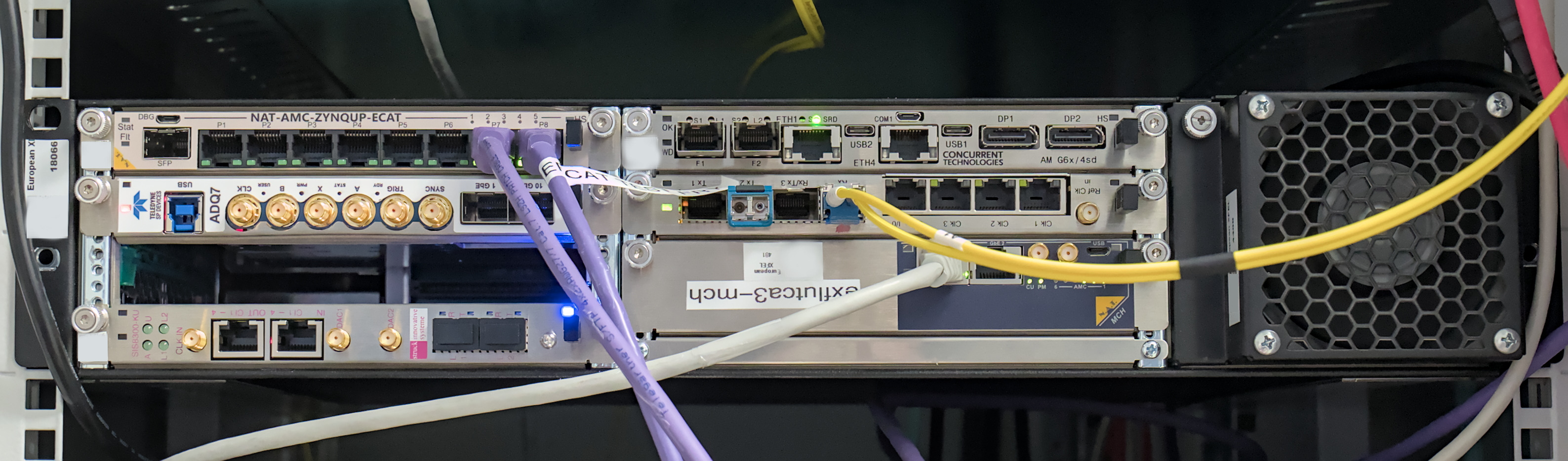}
    \caption{uTCA crate with the timing system (second board in the right column) and the EtherCAT bridge (top board in the left column).}
    \label{fig:utca_crate}
\end{figure}

\begin{figure}
    \includegraphics[width=0.5\textwidth]{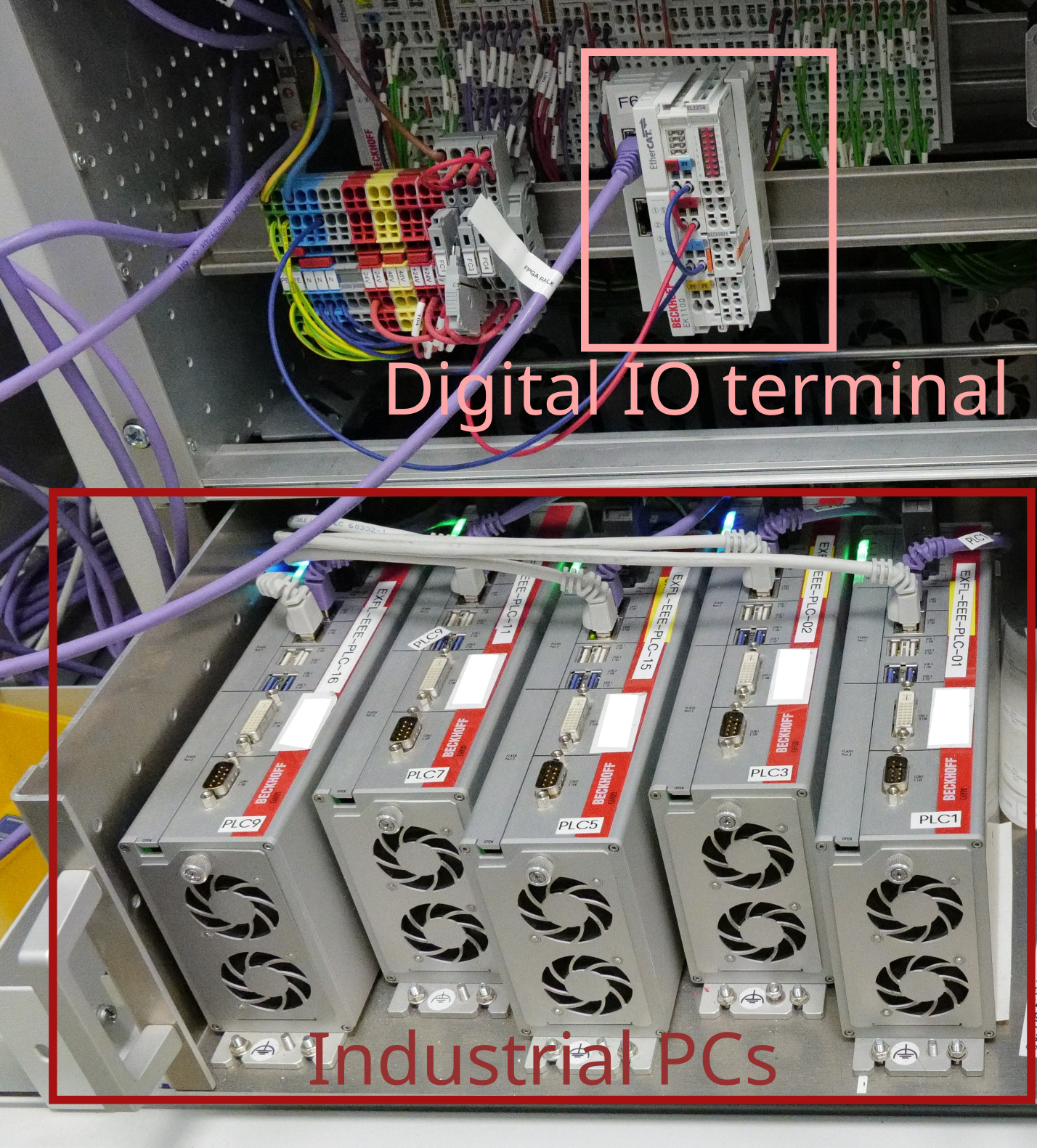}
    \caption{PLC subsystem: industrial PCs (below) and the digital IO terminal EL2258 (upper right).}
    \label{fig:plc_system}
\end{figure}

To test the synchronization between both systems, we built a test setup consisting of a uTCA crate and the PLC components.
Figure\,\ref{fig:utca_crate} shows the uTCA crate.
The crate contains the x2timer board in the second slot of the right column, which receives timing metadata from another x2timer over a yellow optical fiber.
The x2timer distributes the timing metadata over the backplane of the uTCA crate to other boards in the crate.
The EtherCAT bridge board is installed in the uppermost slot in the left column.
Two ports of the board are connected to the PLC master and to the digitial IO terminal EL2258\,\cite{el2258} shown in the figure\,\ref{fig:plc_system}.
The PLC master software is running on an industrial PC under the Windows operating system.
Multiple industrial PCs are visible in the low part of the photo.
The EL2258 terminal is visible in the upper right corner of the photo.

\begin{figure}
    \includegraphics[width=0.5\textwidth]{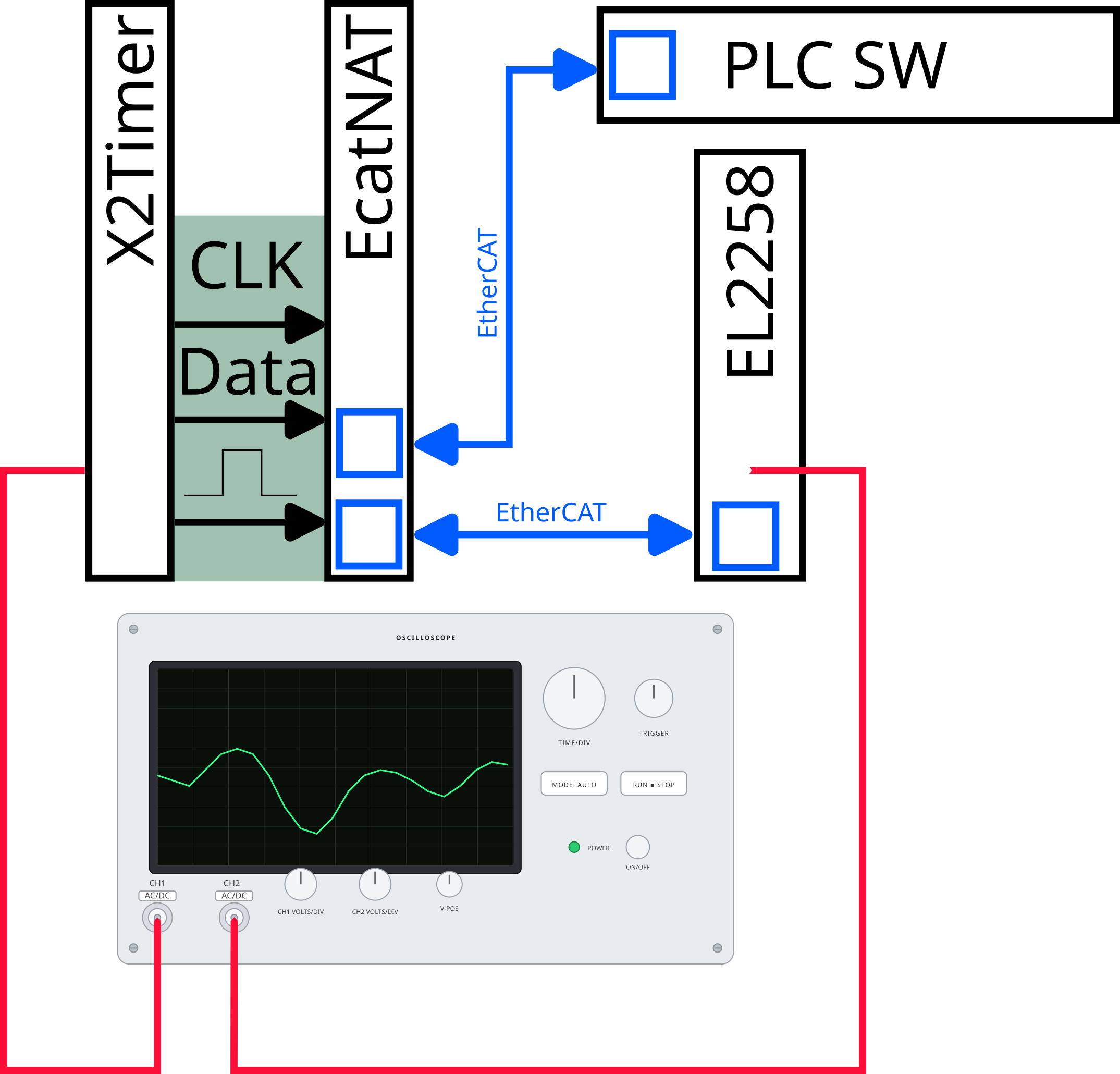}
    \caption{Layout of the setup for testing the synchronization between the timing distribution and the PLC systems.}
    \label{fig:fpga_plc_test_layout}
\end{figure}

Figure\,\ref{fig:fpga_plc_test_layout} shows the layout of the test setup.
The x2timer generates timing metadata and the start-of-cycle pulse, and distributes them to the EtherCAT bridge board over the uTCA backplane.
In addition, the x2timer also provides the start-of-cycle pulse to the input of the oscilloscope LeCroy WR640ZI\,\cite{wr604zi}.
The EtherCAT bridge board latches the arrival time of the start-of-cycle pulse, and makes its timestamp available for the PLC master to read over EtherCAT.
The PLC master software reads the timestamp from the EtherCAT bridge board, adds a constant delay of 6\,ms, and programs the digitial IO terminal to generate a pulse precisely 6\,ms after the start-of-cycle signal.
The digital IO terminal provides the output signal to the oscilloscope which measures the time difference between the start-of-cycle signal and the output of the digital IO terminal.

\begin{figure}
    \includegraphics[width=0.5\textwidth]{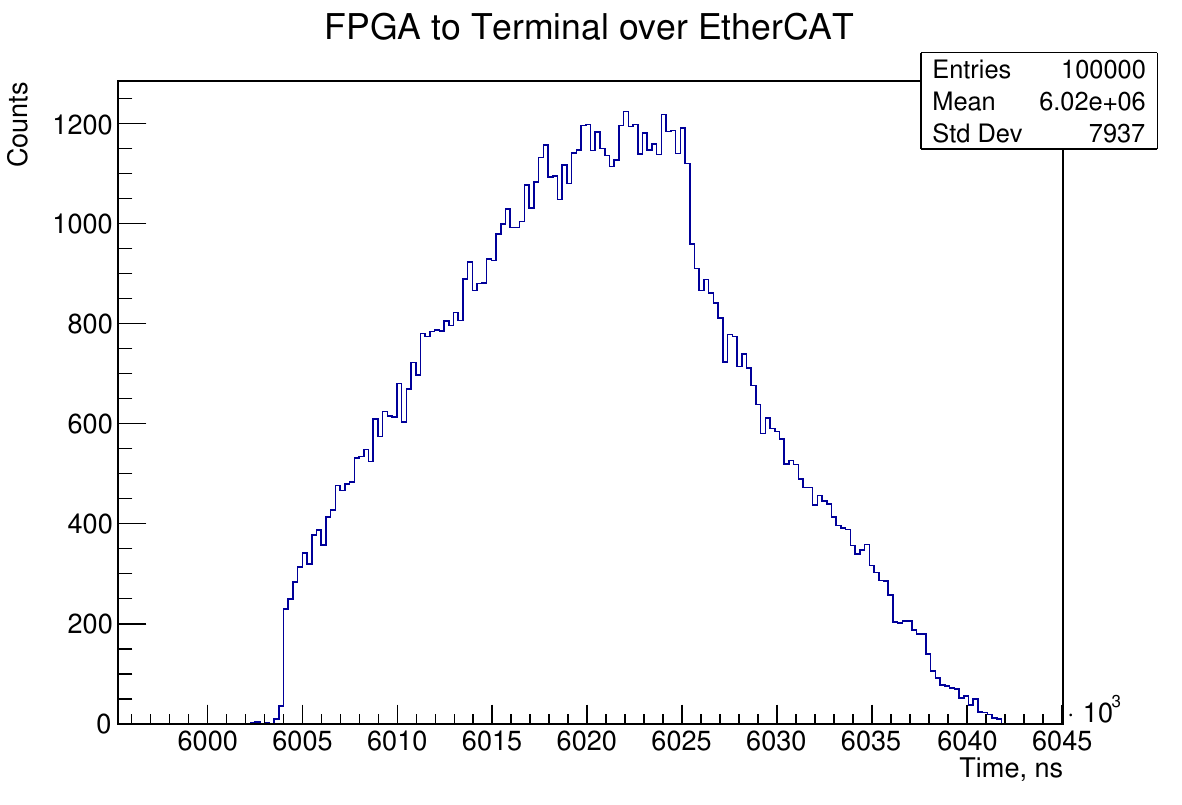}
    \caption{Measured time difference distribution between the trigger signal from the timing system and the pulse generated by the digital IO terminal.}
    \label{fig:fpga_to_plc}
\end{figure}

Figure\,\ref{fig:fpga_to_plc} shows the distribution measured by the oscilloscope.
The peak-to-peak spread of the distribution is 49.8\,µs and the full width at half maximum\,(FWHM) is 20\,µs.
The measurement is consistent with the execution time of the EL2258 terminal of less than 40\,µs\,\cite{el2258}.

\begin{figure}
    \includegraphics[width=0.5\textwidth]{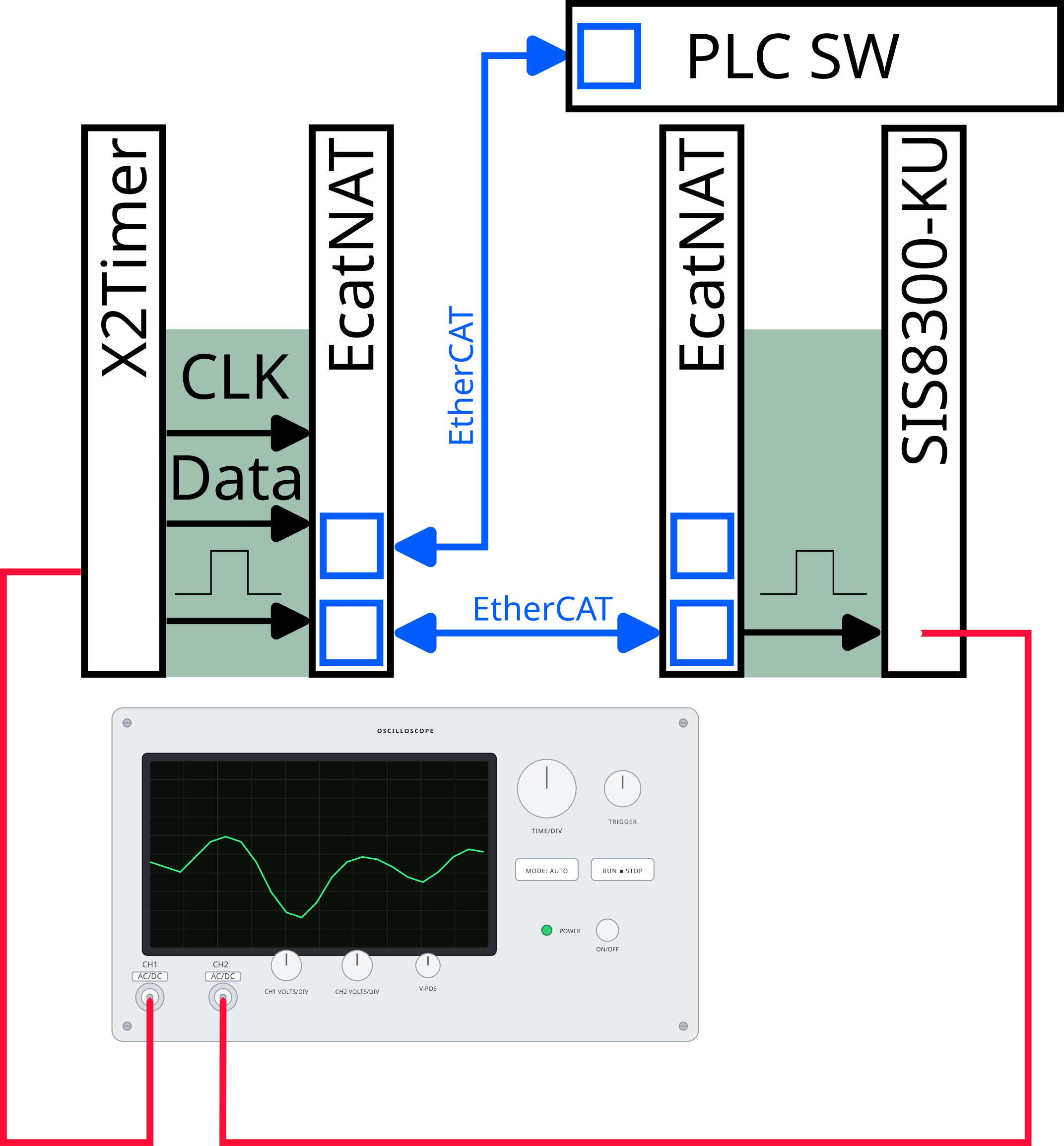}
    \caption{Layout of the setup for testing the synchronization between 2 EtherCAT bridge boards over EtherCAT.}
    \label{fig:fpga_to_fpga_test_layout}
\end{figure}

To eliminate the measurement limitation due to the precision of the signal generation in the terminal, we replaced the terminal with the second EtherCAT bridge board and kept the PLC software algorithm the same.
Figure\,\ref{fig:fpga_to_fpga_test_layout} shows the layout of the updated setup.
Since the EtherCAT bridge board lacks digital outputs from the FPGA, we sent the generated pulse to a SIS8300-KU board over the uTCA backplane.
The FPGA in the SIS8300-KU board asynchronously forwards the signal from the backplane to the front panel digital output, which we connect to the oscilloscope to measure the time difference between the pulses.

\begin{figure}
    \includegraphics[width=0.5\textwidth]{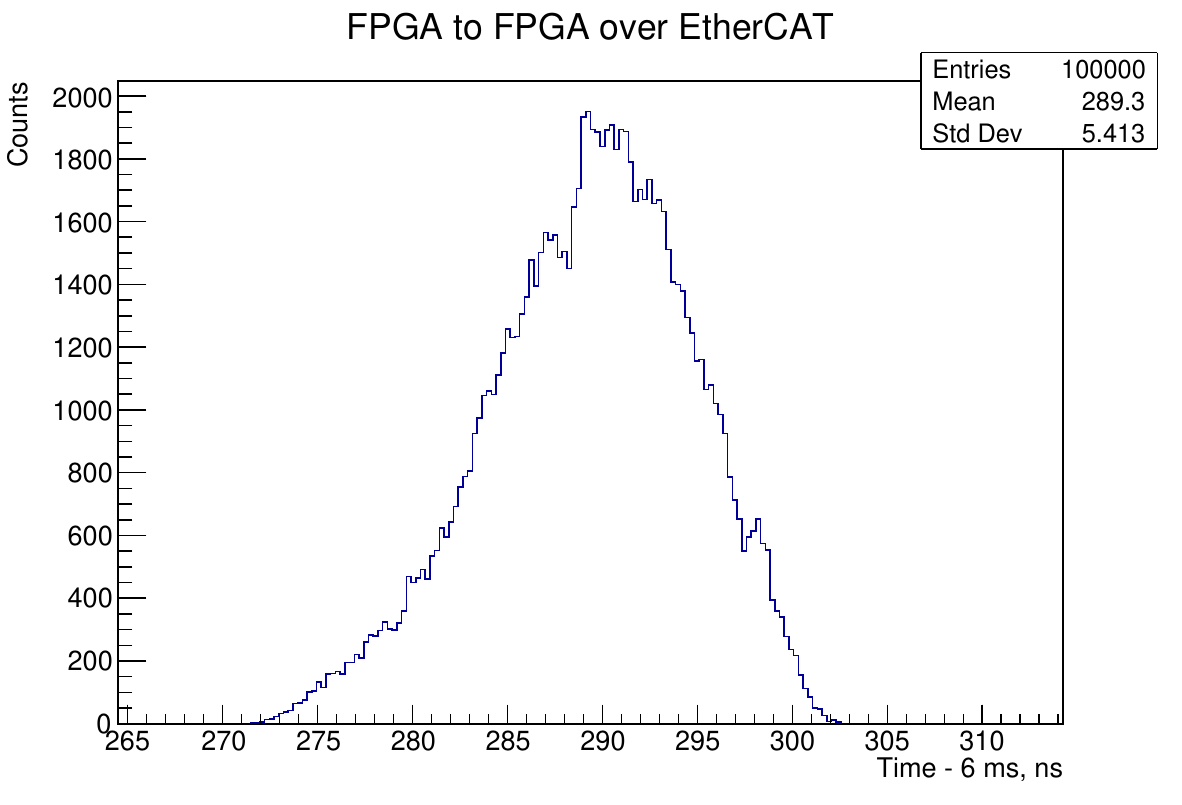}
    \caption{Measured time difference distribution between the trigger signal from the timing system and the pulse generated by the second EtherCAT board, unconnected to the timing system.}
    \label{fig:fpga_to_fpga}
\end{figure}

Figure\,\ref{fig:fpga_to_fpga} shows the distribution acquired with two EtherCAT bridge boards communicating over EtherCAT.
The peak-to-peak spread of the distribution is 49.8\,ns and the FWHM is 12.8\,ns.
This result confirms that the distributed clocks mechanism of the EtherCAT fieldbus can maintain synchronization between endpoints within 50\,ns.
This result is also important because the terminals use a dedicated ASIC to implement EtherCAT slave controllers, whose precision should be compatible with that achieved by FPGA-based endpoints.
The performance degradation which we observed in the first test can be explained by the use of a slower microcontroller in the application layer of the digital IO terminal.

\section{Possible Application of the Synchronization Mechanism at the Femtosecond X-Ray Experiments Instrument}
\label{sec:outlook}

The first application of the proposed solution that we identified is the synchronization of the optical delay line\,(ODL) with the beam at the Femtosecond X-ray Experiments instrument\,(FXE)\,\cite{Galler:xq5006}.

\begin{figure}
    \includegraphics[width=0.5\textwidth]{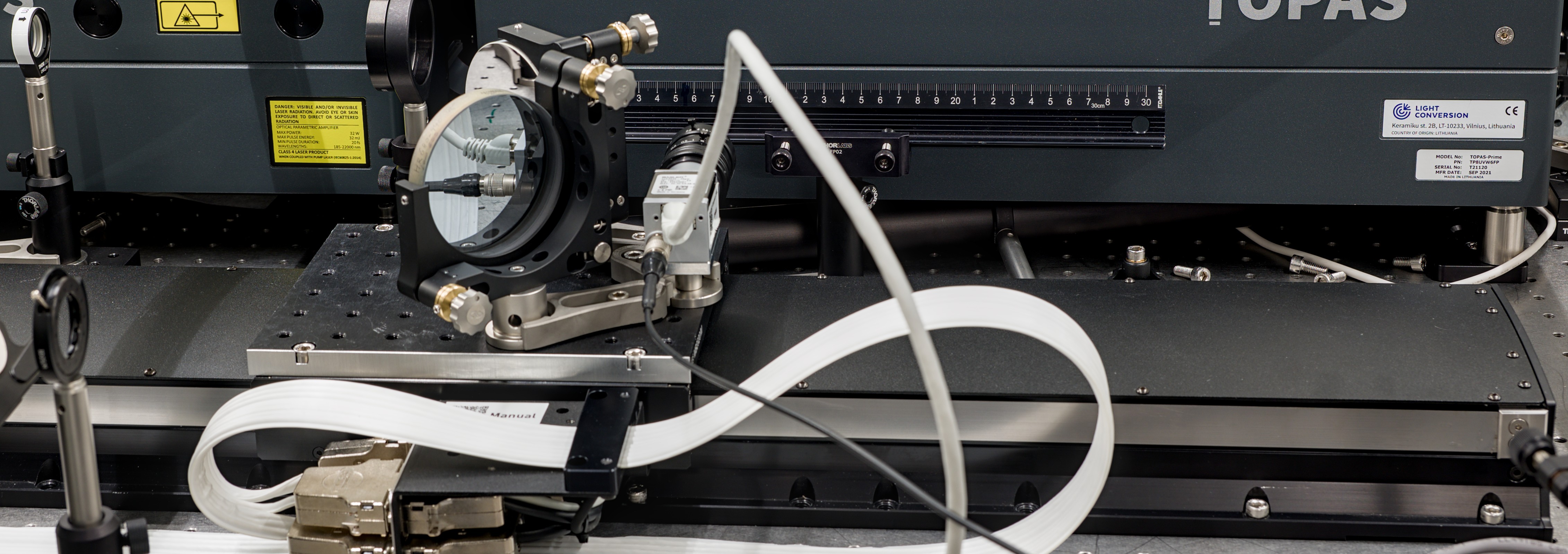}
    \caption{The optical delay line at the FXE instrument. The retroreflector moves along the ODL changing the distance traversed by the optical laser light to change the delay between the optical and the X-ray LASERs in femtosecond range.}
    \label{fig:odl}
\end{figure}

Most experiments at the FXE instrument are carried out to measure the structural evolution of matter during chemical reactions or phase transitions.
This is done by pump-probe measurements, where the delay between two pulses is varied: The first commonly is an optical excitation pulse that initiates a process and the second is a subsequent X-ray probe pulse that measures the atomic or electronic structure.
Both pulses have durations of a few femtoseconds and therefore the delay between the pulses is commonly controlled with few femtosecond accuracy.
Most of the dynamics under investigation are measured over a 100 ps delay range, which corresponds to a 7.5\,mm scan range on an ODL that scales four light paths.
These experiments are often conducted in a stepwise fashion where probe data are accumulated statically at several delay steps.
However, to avoid drift of beam conditions and degradation of the sample during the lengthy delay scans, several user groups have explicitly requested continuous scans of the ODL to accumulate data at all relevant delay positions effectively simultaneously.
This requires moving the ODL with velocities up to 2\,mm/s, so that every second a full delay scan covering 27\,ps delay range is executed.
Accepting a 3\,fs timing jitter due to readout inaccuracies of the ODL corresponds at 2\,mm/s to an acceptable readout timing jitter of 200\,µs.
Specifically at European XFEL, continuously changing the delay implies that the pump-probe delay for the first and last pulse in a burst will be different.
A common burst of 500\,µs will then cause a 13\,fs timing difference from the first to the last pulse when moving at 2\,mm/s, rendering this velocity the upper limit of reasonable motion speeds of an ODL at European XFEL.
Therefore, continuous delay scans at reasonable ODL velocities of 2\,mm/s require a readout timing stability of 200\,µs.

As we have shown in the previous section, the measured precision of the synchronization over EtherCAT is well within the timing stability required by the FXE instrument.
To implement the solution, we will need to select the EtherCAT-based motor encoder with support for the distributed clocks feature and the possibility to schedule the measurement based on the distributed clocks timestamp.

\section{Summary}
\label{sec:summary}

We collaborated with N.A.T. GmbH to build a board in the uTCA form factor that bridges the timing distribution system of the European XFEL and the PLC system using the distributed clocks feature of the EtherCAT fieldbus.
We have implemented the firmware and successfully achieved synchronization between both systems.
By measuring the time difference between signals generated in both domains, we determined the time resolution of the proposed solution to be 49.8\,ns peak-to-peak.
We have also identified a possible application of the device at the FXE instrument which will allow us to improve measurement quality by reducing measurement time.

\Urlmuskip=0mu plus 1mu\relax
\bibliographystyle{IEEEtran}
\bibliography{IEEEabrv,bibliography}

\begin{thebibliography}{10}
\providecommand{\url}[1]{#1}
\csname url@samestyle\endcsname
\providecommand{\newblock}{\relax}
\providecommand{\bibinfo}[2]{#2}
\providecommand{\BIBentrySTDinterwordspacing}{\spaceskip=0pt\relax}
\providecommand{\BIBentryALTinterwordstretchfactor}{4}
\providecommand{\BIBentryALTinterwordspacing}{\spaceskip=\fontdimen2\font plus
\BIBentryALTinterwordstretchfactor\fontdimen3\font minus
  \fontdimen4\font\relax}
\providecommand{\BIBforeignlanguage}[2]{{%
\expandafter\ifx\csname l@#1\endcsname\relax
\typeout{** WARNING: IEEEtran.bst: No hyphenation pattern has been}%
\typeout{** loaded for the language `#1'. Using the pattern for}%
\typeout{** the default language instead.}%
\else
\language=\csname l@#1\endcsname
\fi
#2}}
\providecommand{\BIBdecl}{\relax}
\BIBdecl

\bibitem{euxfel}
T.~Tschentscher, C.~Bressler, J.~Grünert, A.~Madsen, A.~P. Mancuso, M.~Meyer
  \emph{et~al.}, ``{Photon Beam Transport and Scientific Instruments at the
  European XFEL},'' \emph{Applied Sciences}, vol.~7, no.~6, 2017.

\bibitem{kondratenko1980generating}
A.~Kondratenko and E.~Saldin, ``Generating of coherent radiation by a
  relativistic electron beam in an ondulator,'' \emph{Part. Accel.}, vol.~10,
  pp. 207--216, 1980.

\bibitem{ethercat_dc}
{EtherCAT Standardization Committee}, \emph{{{EtherCAT Standard ETG.1000.4, V.
  1.0.4}}}.

\bibitem{plc}
N.~Coppola, J.~Tolkiehn, and C.~Youngman, ``Control using beckhoff distributed
  rail systems at the european xfel,'' \emph{Proceedings of ICALEPCS2013},
  2013.

\bibitem{doi:10.1049/cce:20040104}
D.~Jansen and H.~Buttner, ``Ethernet: Ethercat,'' \emph{Computing and Control
  Engineering}, vol.~15, pp. 16--21, 2004.

\bibitem{ieee8023u1995}
``{IEEE Standards for Local and Metropolitan Area Networks: Supplement to
  Carrier Sense Multiple Access with Collision Detection (CSMA/CD) Access
  Method and Physical Layer Specifications - Media Access Control (MAC)
  Parameters, Physical Layer, Medium Attachment Units, and Repeater for 100
  Mb/s Operation, Type 100BASE-T (Clauses 21-30)},'' Institute of Electrical
  and Electronics Engineers, New York, NY, IEEE Standard IEEE Std 802.3u-1995,
  October 1995, amendment to ISO/IEC 8802-3: 1990.

\bibitem{osi_model}
J.~Day and H.~Zimmermann, ``The osi reference model,'' \emph{Proceedings of the
  IEEE}, vol.~71, no.~12, pp. 1334--1340, 1983.

\bibitem{euxfel_timing}
K.~Rehlich, A.~Aghababyan, H.~Kay, G.~Petrosyan, L.~Petrosyan, V.~Petrosyan
  \emph{et~al.}, ``The new timing system for the european xfel,'' pp.~--, 2013.

\bibitem{x2timer}
A.~Hidvégi, P.~Geßler, H.~Kay, K.~Rehlich, and C.~Bohm, ``{Timing and
  triggering system for the European XFEL project - a double sized AMC
  board},'' in \emph{2012 18th IEEE-NPSS Real Time Conference}, 2012, pp. 1--3.

\bibitem{nat}
{N.A.T. GmbH}. Economic bridge between ethercat and mtca in double-width amc
  form factor.

\bibitem{beckhoff_ipcore}
{Beckhoff Automation GmbH \& Co. KG}, \emph{{{EtherCAT Slave Controller IP Core
  for Xilinx FPGAs Release 3.00k}}}.

\bibitem{systemrdl}
Accellera. {SystemRDL 2.0 Register Description Language}.

\bibitem{peakrdl}
{Alex Mykyta}. {PeakRDL}.

\bibitem{peakrdl-axil-vhdl}
{Dmytro Levit}. {PeakRDL AXI4Lite Master VHDL plugin}.

\bibitem{peakrdl-ethercat-xml}
------. {PeakRDL EtherCAT ESI XML plugin}.

\bibitem{el2258}
{Beckhoff Automation GmbH \& Co. KG}, \emph{{{EL125x, EL2258 8 Channel Digital
  Input/Output Terminal with time stamp. Version: 3.2.0}}}.

\bibitem{wr604zi}
{Teledyne LeCroy Inc.}, \emph{{{Operator's Manual HRO / WaveRunner 6 Zi
  Oscilloscopes}}}.

\bibitem{Galler:xq5006}
A.~Galler, W.~Gawelda, M.~Biednov, C.~Bomer, A.~Britz, S.~Brockhauser
  \emph{et~al.}, ``{Scientific instrument Femtosecond X-ray Experiments (FXE):
  instrumentation and baseline experimental capabilities},'' \emph{Journal of
  Synchrotron Radiation}, vol.~26, no.~5, pp. 1432--1447, Sep 2019.

\end{thebibliography}

\end{document}